\documentclass[preprint, aps,prb,letterpaper,superscriptaddress,showpacs, ]{revtex4}
\usepackage{CJK}
\usepackage{keyval}%
\usepackage{graphicx}
\usepackage{dcolumn}
\usepackage{bm}
\usepackage{color}
\usepackage{amsmath}
\usepackage{soul}

\setkeys{Gin}{width=0.8\columnwidth,clip=true}

\newcommand{\ignore}[1]{}

\begin{document}
\begin{CJK*}{UTF8}{bsmi}
\title{First-principles study on competing phases of silicene: Effect of substrate and strain}
\author{Chi-Cheng Lee (%
李啟正
)}
\affiliation{School of Materials Science, Japan Advanced Institute of Science and Technology (JAIST),
1-1 Asahidai, Nomi, Ishikawa 923-1292, Japan}%
\author{Antoine Fleurence}
\affiliation{School of Materials Science, Japan Advanced Institute of Science and Technology (JAIST),
1-1 Asahidai, Nomi, Ishikawa 923-1292, Japan}%
\author{Rainer Friedlein
}
\affiliation{School of Materials Science, Japan Advanced Institute of Science and Technology (JAIST),
1-1 Asahidai, Nomi, Ishikawa 923-1292, Japan}%
\author{Yukiko Yamada-Takamura
}
\affiliation{School of Materials Science, Japan Advanced Institute of Science and Technology (JAIST),
1-1 Asahidai, Nomi, Ishikawa 923-1292, Japan}%
\author{Taisuke Ozaki
}
\affiliation{School of Materials Science, Japan Advanced Institute of Science and Technology (JAIST),
1-1 Asahidai, Nomi, Ishikawa 923-1292, Japan}%
\affiliation{Research Center for Simulation Science, Japan Advanced Institute of Science and Technology (JAIST),
1-1 Asahidai, Nomi, Ishikawa 923-1292, Japan}

\date{\today}

\begin{abstract}
The stability and electronic structure of competing silicene phases under in-plane compressive stress, either free-standing or on the ZrB$_2$(0001) surface, has been studied by first-principles calculations. A particular ($\sqrt{3}\times\sqrt{3}$)-reconstructed structural modification was found to be stable on the ZrB$_2$(0001) surface under epitaxial conditions. In contrast to the planar and buckled forms of free-standing silicene, in this ``planar-like'' phase, all but one of the Si atoms per hexagon reside in a single plane. While without substrate, for a wide range of strain, this phase is energetically less favorable than the buckled one, it is calculated to represent the ground state on the ZrB$_2$(0001) surface. The atomic positions are found to be determined by the interactions with the nearest neighbor Zr atoms competing with Si-Si bonding interactions provided by the constraint of the honeycomb lattice.
\end{abstract}


\maketitle
\end{CJK*}

\section{Introduction}

Owing to its ability to be either $sp^3$- or $sp^2$-hybridized, under ambient conditions, carbon crystallizes in different forms, such as diamond, 
graphite, graphene, fullerenes and carbon nanotubes. Such diversity does, of course, inspire the search for similar graphitic forms of the element silicon 
which is just below carbon in the periodic table. Whereas silicon nanotubes have already been reported,\cite{Mu,Dmitrii} until recently, 
the two- and three-dimensional graphitic forms of silicon remained elusive but have been anticipated theoretically. 
First-principles studies on the stability and the properties of graphitic Si phases can be traced back decades ago\cite{Cohen} and have mostly focused 
on an atom-thick honeycomb lattice of Si atoms, called silicene, as the Si-counterpart of graphene. In contrast to graphene, however, 
it is well accepted that silicene is unstable as long as the structure is planar.\cite{Cohen,Takeda,Ciraci} Instead, free-standing silicene 
is predicted to be stable in a so-called ``buckled'' structure, where the two sub-lattices of the bipartite lattice are at different heights.\cite{Ciraci} 
Like in graphene, charge carriers in free-standing silicene are found to be massless Dirac fermions even if it is buckled.\cite{Takeda,Ciraci} 
The Fermi velocity is calculated to be in the order of $10^6 m/s$.\cite{Ciraci} Additionally, due to a spin-orbit coupling much stronger than that in graphene, 
the quantum Hall effect is predicted to occur\cite{Ciraci,Liu} and, owing to its buckled structure, to be switchable by the application of a perpendicular electric field.\cite{Ezawa} 
Since the many fascinating properties and countless potential applications proposed for graphene\cite{Fujita,Novoselov,Zhang,Barone,Son,Berger,Katsnelson,Geim,Han,Li,Wang} 
may occur in silicene as well, it is of great importance to uncover all stable structural modifications of silicene. 

So far, Si honeycomb structures have been reported to grow on a limited number of substrates, 
such as Er, Ag, ZrB$_2$, and Ir.\cite{Veuillen,Wetzel,Kawai,Feng,Jamgotchian,Daniele,Vogt,Chen,Chen2,Fleurence,Meng} DFT calculations 
in conjunction with experimental observations point out that all of those two-dimensional forms of silicon are deviating 
from the expected buckled form of free-standing silicene.\cite{Kawai,Feng,Jamgotchian,Daniele,Vogt,Chen,Chen2,Fleurence,Meng} 
Among them, those on Ag(111) and on ZrB$_2$(0001) surfaces are the most exciting ones 
since the experimental observation of $\pi$ bands links them to the theoretical concept of silicene. Imposed by epitaxial conditions, 
the type and degree of buckling can vary by a large amount which in turn influences the electronic properties as well. 
The intimate relationship between the structure and the electronic properties makes the atomic-scale buckling 
a highly relevant parameter in the description of silicene. On the other hand, the structural flexibility, related to a mixed $sp^2/sp^3$ hybridization, 
may allow the engineering of desired properties 
such as the opening of a gap, something that is difficult to achieve in purely $sp^2$-hybridized, robust graphene.\cite{Fleurence} 

For silicene on the Ag(111) surface, the growth behavior is strongly dependent on the temperature, the coverage and the deposition rate.\cite{Kawai,Feng,Jamgotchian,Daniele,Vogt,Chen} 
Among the reported superstructures, post-annealing cannot switch one observed phase to another\cite{Jamgotchian} which advocates the presence of non-negligible energy barriers 
between these phases. It has been proposed that all the observed silicene superstructures exhibit different orientations with respect to the silver substrate 
which is manifested in particular surface reconstructions.\cite{Chen2} On the contrary, epitaxial silicene on zirconium diboride (0001), 
formed by the segregation of atoms from the substrate on the surface of ZrB$_2$(0001) thin films grown on Si(111) wafers,\cite{Fleurence} has the advantage to exhibit a single orientation, 
to cover the whole sample surface homogeneously, and to be very well reproducible. The latter is a direct consequence of the spontaneous and self-terminating growth mode. 
Epitaxial silicene on ZrB$_2$(0001) is ($\sqrt{3}\times\sqrt{3}$)-reconstructed which is provided by the commensurate relationship of the ($\sqrt{3}\times\sqrt{3}$) unit cell of silicene 
with the (2$\times$2) unit cell of ZrB$_2$(0001).\cite{Fleurence} The spontaneous formation of silicene on ZrB$_2$(0001) suggests that this particular reconstruction is important for the stability of silicene. 
The nature of this ($\sqrt{3}\times\sqrt{3}$) reconstruction has recently been discussed based on density functional theory (DFT) calculations.\cite{Fleurence} Note, however, 
that the considered phase, whose calculated structural and electronic properties are in partial agreement with the experimental findings,\cite{Fleurence, Friedlein} 
has been calculated to be a metastable structure. It deviates from that of the predicted free-standing silicene by the position of one of the atoms forming the bottom sub-lattice lifted 
up to almost the height of the top sub-lattice.\cite{Fleurence} On the other hand, the properties of the ground state that are discussed in the present paper for the first time 
do not fit with the experimental results. In the ground state proposed by DFT, the structure is ``planar-like'', closer to that of planar graphene than to the metastable phase. 
In this structure, five out of the six Si atoms per hexagon reside in a plane characterized by a residual buckling of just 0.01~\AA~, with the remaining atom protruding at a height of 1.58~\AA. 

Note that as a similar ($\sqrt{3}\times\sqrt{3}$)-reconstructed planar-like phase has been calculated to be one of the silicene structures forming on the Ag(111) surface,\cite{Chen2} 
it is of great interest to investigate under which conditions this particularly reconstructed phase may become stable. Here, interactions with the substrate and epitaxial strain are important issues. 
In this paper, we report the results of our theoretical study describing the competition between structural variations of ($\sqrt{3}\times\sqrt{3}$)-reconstructed silicene under the influence of the ZrB$_2$(0001) 
substrate and in-plane compressive strain. The discussion sheds light onto both the structural parameters and the role of the substrate electronic properties on the stability of epitaxial silicene. 
The paper is organized as follow: After presenting the details of the computational methods (section II), we will discuss the stability of the different silicene structures in a stepwise manner. 
First, free-standing silicene (section III) and the ZrB$_2$(0001) substrate itself (section IV) are considered. In section V and section VI are presented 
the interactions of single Si atoms and of Si honeycomb layers in contact with the ZrB$_2$(0001) substrate, respectively. Discussion and conclusions can be found in section VII.

\section{Computational detail}

Density Functional Theory (DFT) calculations within a generalized gradient approximation\cite{Kohn,Perdew} have been performed using the OpenMX code which is based on norm-conserving pseudopotentials 
generated with multi reference energies\cite{MBK} and optimized pseudoatomic basis functions.\cite{Ozaki,openmx} The regular mesh of 270 Ry in real space was used for the numerical integrations and for the solution of the Poisson equation. The cut-off radius of 7 Bohr has been chosen 
for all the basis functions. For each Zr atom, three, three, and two optimized radial functions have been allocated for the $s$-, $p$-, and $d$-orbitals, respectively, 
as denoted by Zr-$s3p3d2$. For B and Si atoms, B-$s4p2d1$ and Si-$s2p2d1$ configurations have been adopted. The spin-orbit coupling has not been considered in our calculation. 
The calculated lattice constants of bulk ZrB$_2$, $a$ = 3.17~\AA~ and $c$ = 3.55~\AA, obtained using a 8$\times$8$\times$5 mesh of $k$ points in reciprocal space are in good agreement 
with experimental data.\cite{Vajeeston} For the study of free-standing silicene, of the ZrB$_2$(0001) surface and of silicene on the ZrB$_2$(0001) surface, 
the in-plane lattice constants of the bulk and a 4$\times$4$\times$1 mesh of $k$ points have been used for the unit cell corresponding to the (2$\times$2)-reconstructed ZrB$_2$(0001) surface. 
For the calculation of surface properties, the structural parameters of the central two Zr and three B layers of a 15-layer Zr-terminated slab have been fixed to the bulk values. 
Other atomic positions have been relaxed until the residual force on each atom has reached values of less than $3\times10^{-4}$ Hartree/Bohr. In order to allow for a symmetric arrangement, 
silicene layers containing six Si atoms per (2$\times$2) unit cell of the ZrB$_2$(0001) surface have been placed on both sides of the Zr-terminated slab. 

In terms of the representation of the electronic band structures, the Brillouin zones of the (1$\times$1) unit cell of silicene and the (1$\times$1$\times$1) unit cell of bulk ZrB$_2$ have been chosen for the studies of free-standing silicene and ZrB$_2$(0001) surface, respectively. 
For the case of silicene on the ZrB$_2$(0001) surface, both Brillouin zones have been chosen. In the study of silicene-derived bands, the two symmetric silicene layers on each surface of the slab 
are considered as two separate layers. Since the supercell is larger than the primitive unit cell chosen for the representation, 
an additional procedure\cite{Wei} to unfold the bands into larger zones is needed and has been generalized to the non-orthogonal pseudoatomic basis functions.\cite{Chi} 
In order to show the spectral weight related to each eigenstate of the system at a $k$ point, we used the so-called ``fat band representation'', where the weight was represented by the diameter of a circle. The respective contributions from different orbitals to the spectral weight are given by the diameters of color-coded circles, as specified in the figure captions. The diameter of grey-colored circle is obtained by summing over all the orbital contributions of an eigenstate. For a better visualization, eigenstates with the same $k$ and separated by an energy differences less than 0.02 eV are considered as being degenerated and represented by a single circle.

In order to assure structural convergence of the ground state of silicene on the ZrB$_2$(0001) surface calculated by DFT, additionally, we have performed molecular dynamics calculations 
starting from different initial structures, the temperature set to 400 K. The same ground-state structure has been found in all of the molecular dynamics simulations. 
A consistent ground-state structure of silicene on the (2$\times$2) unit cell of the ZrB$_2$(0001) surface was also found by using the VASP package with ultrasoft pseudopotentials.\cite{Kresse,Vanderbilt} 
To analyze the energy barrier between the ground state and a proposed metastable phase, 16 transition images are adopted using the nudged elastic band (NEB) method where the minimum energy path between two configurations can be found.\cite{Jonsson} 
Here, the maximum force is less than $5\times10^{-3}$ Hartree/Bohr.

\section{Free-standing silicene}

\begin{figure}[p]
\includegraphics[width=0.4\columnwidth,clip=true,angle=90]{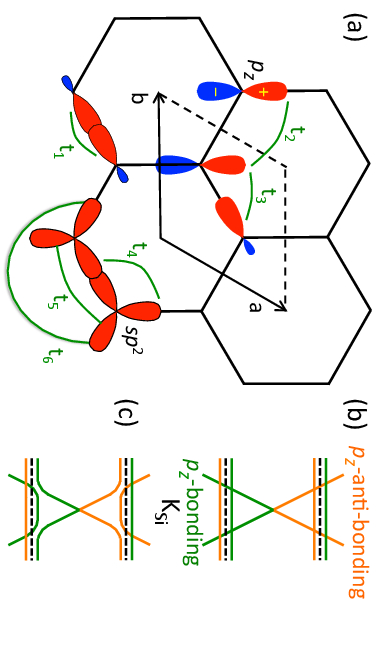}
\caption{\label{fig:fig0}
(a) Nearest neighbor hopping integrals within a simple tight-binding model for a Si honeycomb lattice considering only the  $\sigma$-bond hopping ($t_1$), $p_z-p_z$ hopping ($t_2$), 
and the dominant $p_z-sp^2$ hopping ($t_3$). $t_3$ is zero for the planar structure but can reflect the strength of buckling between the two atoms in the unit cell. 
A richer dispersion in the $sp^2$ branches can further be introduced by including other hopping terms, by for example, $t_4$, $t_5$, and $t_6$. (b) Scheme of the band structure 
revealed by applying the Hamiltonian in Eq.~(\ref{eq:eqn1}) to the $t_1$ and $t_2$ terms alone. 
Horizontal lines represent $\sigma$ bands and the X-shape lines illustrate the $\pi$ bands forming a Dirac-cone structure at $K$. 
The bonding and antibonding characters of bands in the $p_z$ sub-space are indicated by green and orange colors, respectively. The black dashed lines show bands without $p_z$ character. (c) Schematic band structure with non-zero $t_3$.
}
\end{figure}

Determined by the $sp^2$ hybridization similar to graphene, free-standing planar silicene is expected to display comparable electronic properties.\cite{Ciraci} As shown in Fig.~\ref{fig:fig0} (a), 
several hopping parameters can contribute to the electronic band structure of planar silicene. This can well be explained within a tight-binding approach considered here initially. 
A finite set of hopping integrals related to the $\sigma$-bond hopping between two of the $sp^2$ orbitals ($t_1$), the $p_z-p_z$ hopping ($t_2$), and the dominant $p_z-sp^2$ hopping ($t_3$) 
shall be used. In reciprocal space, the Hamiltonian, $H(\vec{k}=k_a \vec{a}^* + k_b \vec{b}^*)$, is given by Eq.~(\ref{eq:eqn1}). Here, $\vec{a}^*$ and $\vec{b}^*$ 
denote the reciprocal lattice vectors of the unit cell containing two Si atoms. As a consequence of the given symmetry, this Hamiltonian describes several interesting properties of the planar structure 
of silicene which are discussed below. 

Because of the planarity, $t_3$ is zero. Therefore, $H(\vec{k})$ in the $p_z$ sub-space is decoupled from that of the other orbitals. As a consequence, this simply leads to the presence 
of a bonding ($\pi$) band and an anti-bonding ($\pi^{*}$) band. The off-diagonal term ($t_2 + t_2 e^{i 2\pi k_a} + t_2 e^{-i 2\pi k_b}$) becomes zero at $\vec{k}=\vec{K}\equiv(\frac{1}{3},\frac{1}{3})$ 
forming a Dirac point, and a linear dispersion can be derived as $E(\vec{k})=\epsilon_{p_z}\pm \sqrt{3} |t_2| a q / 2$ in the vicinity of the $K$ point, where $\vec{q}=\vec{k}-\vec{K}$. 
This is illustrated in Fig.~\ref{fig:fig0} (b). In the half-filled case, the Fermi energy is equal to the site energy $\epsilon_{p_z}$ and thus coincides with the energy of the Dirac point. 
The simple analysis reveals the unique signature of the $p_z$-derived bands formed in the honeycomb lattice. On the other hand, within this model, the $sp^2$ orbitals form flat bands 
at energies of $\epsilon_{sp^2} \pm |t_1|$. 

Evolving from the planar to the buckled structure, the hopping integral $t_3$ which reflects the strength of the buckling becomes non-zero. 
As indicated in Eq.~(\ref{eq:eqn1}), the orbitals are then coupled to each other with different phases. For planar silicene, 
the degenerate eigenstates at the Dirac point can be viewed as two independent Bloch states. Each of them involves only Si $p_z$ orbitals of one of the two sub-lattices 
and must show a double degeneracy (if the spin is not taken into account). In fact, the buckling cannot lift the degeneracy of the two states. 
For example, one can easily rotate the eigenstates at the Dirac point to allow each eigenstate to couple only with one of the Si $p_z$ orbitals such that the degeneracy 
is still preserved by a non-zero $t_3$. The absence of a gap opening at the Dirac point indicates that the buckling is not associated with a gain in band energy. 

\begin{widetext}
\begin{eqnarray}
\label{eq:eqn1}
\begin{pmatrix} \epsilon_{sp^2} & 0 & 0 & 0 & t_1 & 0 & 0 & t_3 \\
                0 & \epsilon_{sp^2} & 0 & 0 & 0 & t_1 e^{i 2\pi k_b} & 0 & t_3 e^{i 2\pi k_b} \\
                0 & 0 & \epsilon_{sp^2} & 0 & 0 & 0 & t_1 e^{-i 2\pi k_a} & t_3 e^{-i 2\pi k_a} \\
                0 & 0 & 0 & \epsilon_{p_z} & -t_3 & -t_3 e^{i 2\pi k_b} & -t_3 e^{-i 2\pi k_a} & t_2 + t_2 e^{-i 2\pi k_a} + t_2 e^{i 2\pi k_b}\\
                t_1 & 0 & 0 & -t_3 & \epsilon_{sp^2} & 0 & 0 & 0 \\
                0 & t_1 e^{-i 2\pi k_b} & 0 & -t_3 e^{-i 2\pi k_b} & 0 & \epsilon_{sp^2} & 0 & 0 \\
                0 & 0 & t_1 e^{i 2\pi k_a} & -t_3 e^{i 2\pi k_a} & 0 & 0 & \epsilon_{sp^2} & 0 \\
                t_3 & t_3 e^{-i 2\pi k_b} & t_3 e^{i 2\pi k_a} & t_2 + t_2 e^{i 2\pi k_a} + t_2 e^{-i 2\pi k_b} & 0 & 0 & 0 & \epsilon_{p_z} \\
\end{pmatrix}
\end{eqnarray}
\end{widetext}

Another interesting finding revealed by the analysis of $H(\vec{k})$ is related to the bonding state characterized by $E=\epsilon_{sp^2}-|t_1|$ and 
the corresponding anti-bonding state with $E=\epsilon_{sp^2}+|t_1|$. These states are preserved even in the case of a non-zero $t_3$ and carry no $p_z$ component. 
The energies of the two states are immune to the buckling. Within the whole Brillouin zone, the corresponding eigenstates involve only $sp^2$ orbitals. Unlike these two flat bands, 
all the other bands derive from a mixture of $sp^2$ and $p_z$ orbitals. The three bonding bands and three anti-bonding bands defined in the $p_z$ sub-space are illustrated in Fig.~\ref{fig:fig0} (b). 
The buckling gives rise to an additional modification in these electronic states manifested in an avoided crossing of the bonding-type (or anti-bonding-type) bands as shown in Fig.~\ref{fig:fig0} (c). 

\begin{figure}[p]
\includegraphics[width=0.70\columnwidth,clip=true,angle=90]{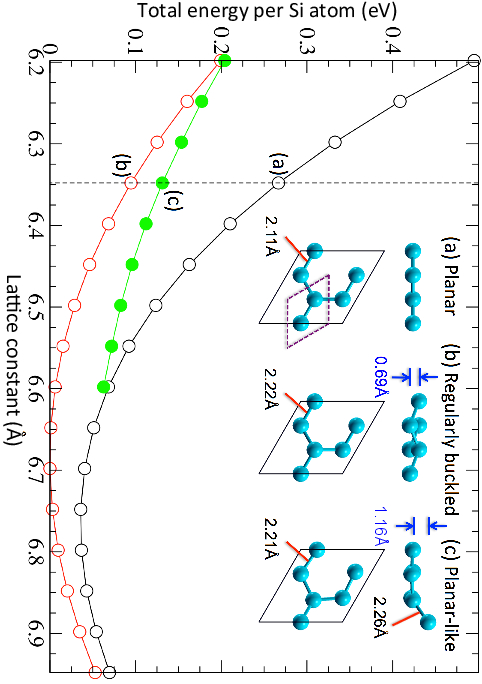}
\caption{\label{fig:fig1} 
Free-standing (a) planar, (b) regularly-buckled and (c) planar-like structures of silicene and their corresponding total energies as a function of the lattice constant $a$ 
of the ($\sqrt{3}\times\sqrt{3}$) unit cell of silicene. The bond lengths and the buckling heights of the three phases are indicated for the particular lattice constant of $a$ = 6.35~\AA~ which corresponds to that 
of the (2$\times$2) unit cell of the ZrB$_2$(0001) surface. This particular value of the lattice constant is indicated by the dashed line. 
Upon increase of $a$ to more than 6.60~\AA, the planar-like structure relaxes to the planar structure. 
}
\end{figure}

In the framework of the DFT calculations, in addition to the planar and buckled structures of silicene, we considered the ($\sqrt{3}\times\sqrt{3}$)-reconstructed structure 
as well since it has been predicted to be relevant for epitaxial systems.\cite{Fleurence,Chen2} For comparison, we also evaluated the planar and buckled forms within a supercell in the size of the primitive unit cell of this ``planar-like'' structure. Hereafter, the buckled phase will be referred to as the ``regularly-buckled'' one in order to distinguish the two non-planar phases. In Fig.~\ref{fig:fig1}, the geometrical structures of the planar, the regularly-buckled, and the planar-like phases are plotted together 
with their total energies as a function of the in-plane lattice constant $a$ of the ($\sqrt{3}\times\sqrt{3}$) unit cell. 

With one Si atom per unit cell protruding out of plane, the planar-like phase is able to sustain a longer in-plane bond length as compared to the planar phase 
such that it becomes closer to that of the regularly-buckled one. As shown in Fig.~\ref{fig:fig1}, in regularly-buckled silicene, 
the buckling height of 0.69~\AA~ at $a$ = 6.35~\AA~ decreases to 0.50~\AA~ at $a$ = 6.70~\AA~ which is a result of a tendency 
to maintain the bond length for a large range of in-plane strain. For $a$ = 6.20~\AA~ to $a$ = 6.60~\AA, which is close to the value provided by ZrB$_2$ as a substrate,
the total energy of planar-like silicene is in between the planar and regularly-buckled forms. The relationship between the total energy and the lattice constant suggests 
that under larger in-plane compressive stress, the planar-like phase might take over the regularly buckled phase.

\begin{figure}[p]
\includegraphics[width=0.80\columnwidth,clip=true,angle=0]{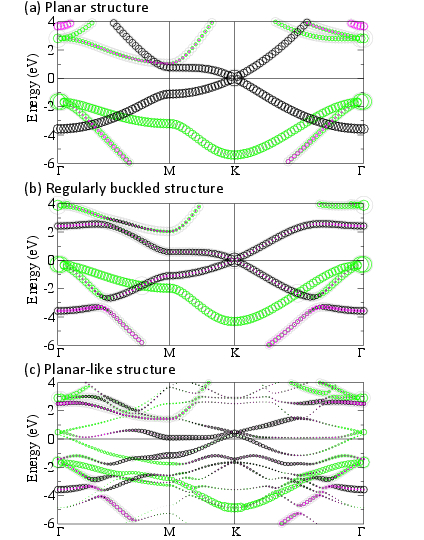}
\caption{\label{fig:fig2} 
The electronic band structures of free-standing (a) planar, (b) regularly-buckled, and (c) planar-like phases 
as unfolded from the ($\sqrt{3}\times\sqrt{3}$) unit cell ($a$ = 6.35~\AA, using the in-plane lattice constants of the (2$\times$2) unit cell of ZrB$_2$(0001)) 
into the first Brillouin zone of (1x1) silicene. 
The $s$ and $p_z$ characters of bands are colored in magenta and black, respectively, and that of $p_x$ and $p_y$ orbitals in green. 
The chemical potentials have been set to zero. The difference in chemical potentials was smaller than 0.3 eV. 
}
\end{figure}

Since the total energies of regularly-buckled and planar-like phases are generally lower than that of the planar one, it is important to verify again if the energy gain might be of electronic origin or not. 
For this purpose, we consider the DFT results for $a$ = 6.35~\AA~ which is the in-plane lattice constant of the (2$\times$2) unit cell of ZrB$_2$(0001). The electronic structures of these three phases 
represented in the first Brillouin zone of the silicene (1$\times$1) unit cell are shown in Fig.~\ref{fig:fig2}. As expected, the band structures of the planar and regularly-buckled phases are quite similar, 
and do not show any additional symmetry breaking introduced by the ($\sqrt{3}\times\sqrt{3}$) supercell. In Figs.~\ref{fig:fig2} (a) and (b), 
occupied $\pi$ and unoccupied $\pi^{*}$ bands with $p_z$ character are colored in black. At the $K$ point, a Dirac cone is found at $E_F$ which is in agreement 
with our analysis using the simple tight-binding model and earlier studies.\cite{Takeda,Ciraci}

While for planar silicene, the $p_z$ orbitals do not couple to the $sp^2$ orbitals, the buckling leads to non-zero hopping integrals between the $p_z$ 
and the neighboring $sp^2$ orbitals which confer a mixed $\pi-\sigma$ character to the originally pure $\pi$ and $\pi^{*}$ bands. 
Unlike one of the $\sigma$ band marked in green in Figs.~\ref{fig:fig2} (a) and (b) that carries negligible $p_z$ weight, 
the other occupied $\sigma$ band mixes with the bonding $\pi$ band such that band crossing is avoided. Although the resulting gap is large, 
the corresponding bonding and anti-bonding bands are both occupied such that no band energy is gained. Similarly, the energy of the $\sigma$ band at the $\Gamma$ point 
is increased by more than 1 eV. Therefore, the mechanism which stabilizes the regularly-buckled silicene over the planar one is hard to visualize 
in the electronic band structure itself. It may instead relate to an instability in the phonon part that involves the lattice repulsive potential 
and the response of electrons to the lattice vibration.\cite{Takeda,Ciraci}   

The electronic structure of planar-like silicene is presented in Fig.~\ref{fig:fig2} (c). A strong breaking of the symmetry causes 
back-folding and the lifting of the degeneracy of bands. Despite the back-folding, some resemblance to planar silicene is found. 
In particular, the energies of the $\sigma$ band at the $\Gamma$ point are similar. 
If the origin of the stabilization of the planar-like phase with respect to the planar one in terms of the total energy is not found in the band structure, 
other energy terms shall be considered. And indeed, it is found that energy is gained mainly from a reduction of the core-core Coulomb repulsion of the nuclei. 
Quantitatively, this repulsion is lowered by about 50 eV and 60 eV per Si atom, as compared to the regularly-buckled and planar phases, respectively. 
This suggests an essential role of the atomic positions and the bond lengths for the stability of the planar-like phase that might provide a competitive advantage 
over other phases under restraints imposed by epitaxial conditions on substrates. Note that cone-like band dispersions are still noticeable even 
under the ($\sqrt{3}\times\sqrt{3}$) reconstruction which, however, leads to the opening of a small gap and an up-shift of the corresponding features. A similar dispersion was also found in the band structure calculated for freestanding silicene having the structure of epitaxial silicene on Ag(111) substrate.\cite{Chen2} 
This can be visualized at the $K$ point via unfolding, as shown in Fig.~\ref{fig:fig2}. 

Considering again a tight-binding approach using Wannier functions,\cite{Marzari,Weng} the deviation of the band structure of the planar-like phase from those of both the planar and regularly-buckled phases can be explained by the breaking of both the three-fold rotation and inversion symmetries. This is also expressed in the splitting of on-site energies and hopping integrals provided in Table~\ref{tab:hoppings}. Note that for the planar-like phase, not only differences between the two sub-lattices but also between individual atoms can be identified. 

\begin{table}[h]
\caption{\label{tab:hoppings}
Hopping integrals of free-standing planar, regularly-buckled, planar-like phases at $a$ = 6.35~\AA.
On-site energy is denoted by $\epsilon$ and the $t_1$, $t_2$, and $t_3$ are defined in Fig.~\ref{fig:fig0} (a). For the planar-like structure,
the protruding Si atom is denoted by Si C. Si B is defined as one of the atoms belonging to the sub-lattice without the protruding Si atom.
Si A is defined as one of the two other Si atoms belonging to the sub-lattice of the protruding Si atom. The number in the parenthesis is the one associated with the protruding Si atom. The unit is eV.
}
\begin{ruledtabular}
\begin{tabular}{cccccc}
          &   $\epsilon_{sp^2}$ & $\epsilon_{p_z}$ &  $t_1$   &       $t_2$   &         $t_3$  \\
\hline
planar    &    0.18             &    0.38          &  -6.76        &  -1.56   &         0.00    \\
\hline
regularly-buckled   &    0.30             &    0.82          &  -5.61        &  -0.90   &         1.57, -1.57  \\
\hline
planar-like \\
Si A      &   -0.51             &    0.54          &  -6.05        &  -1.47          &  0.35, -0.33   \\
Si B      &    0.30 (-0.36)     &    0.65          &  -6.05 (-4.68) &  -1.47 (-0.23) &   0.35, -0.33 (2.08, -2.64)    \\
Si C      &   -0.72             &   -0.22          &  -4.68        &  -0.23          &  2.08, -2.64   \\
\end{tabular}
\end{ruledtabular}
\end{table}

\begin{figure}[p]
\includegraphics*[width=0.7\columnwidth,clip=true,angle=90]{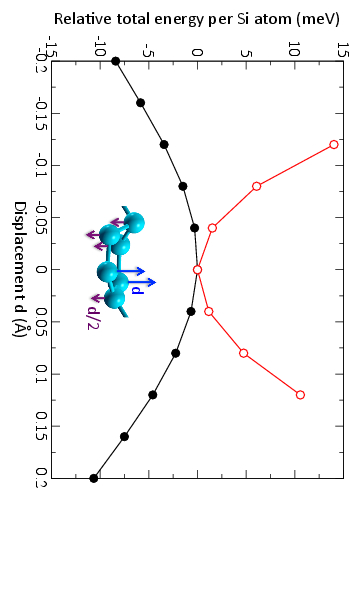}
\caption{\label{fig:figvib}
Variation of the total energy as a function of the atomic displacement $d$ related to the indicated vibration, 
for the free-standing planar-like phase (solid circles) and the planar-like silicene  
on the ZrB$_2$(0001) surface (open circles). The energies of extrema are set to zero.
}
\end{figure}

Finally, we address the stability of planar-like silicene related to the out-of-plane displacement of some of the atoms. 
In Fig.~\ref{fig:figvib} is shown the total energy as a function of the displacement of the two other Si atoms belonging to the sub-lattice of the protruding Si atom 
against the remaining four Si atoms of the reconstructed unit cell. 
The displacement with an equal amount for both atoms is associated with the $\Gamma$-point vibration that can lead to two structural modifications with a different type of buckling. 
As shown in Fig.~\ref{fig:figvib}, the energy of the planar-like structure becomes lower when the atoms of free-standing silicene are displaced. However, such an instability might disappear under certain circumstances, 
like for example, in the presence of a substrate. With the presentation of the total energy as a function of the amount of displacements, for the case of planar-like silicene relaxed onto the surface of ZrB$_2$(0001), a stable Born-Oppenheimer surface can be found. The dramatical change of the total energy surface implies that the substrate does play a crucial role in the stabilization of epitaxial silicene structures. 

\section{The zirconium diboride (0001) surface}

\begin{figure}[p]
\includegraphics[width=0.60\columnwidth,clip=true,angle=90]{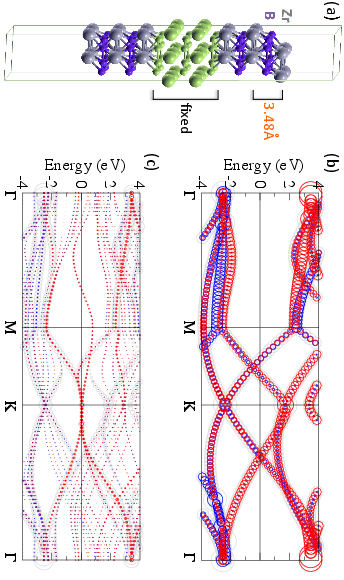}
\caption{\label{fig:fig3}
(a) Structure of the slab used in the calculations of properties of the (2$\times$2) unit cell of the Zr-terminated ZrB$_2$(0001) surface. The central five layers have been fixed to the bulk structure. 
The respective electronic band structures of (b) bulk ZrB$_2$ and (c) the slab, both represented within the first Brillouin zone of bulk ZrB$_2$. The Zr $d$ and B $p$ 
characters of bands are colored in red and blue, respectively. Only the contribution of the outermost ZrB$_2$ layers of the slab are shown in the fat-band representation. 
Newly revealed bands in the vicinity of the Fermi energy are identified by their dispersions as the surface states. 
}
\end{figure}

Before going to epitaxial silicene on the diboride substrate, the electronic structure of the bare Zr-terminated ZrB$_2$(0001) surface shall be discussed. 
Electronic structure calculations with results similar to those published previously  \cite{Otani,Aizawa,Kumashiro,Chi} have been performed for the bulk and for a slab, 
the model of which is shown in Fig.~\ref{fig:fig3} (a). Electronic states related to contributions 
from orbitals of the bulk and of the outermost ZrB$_2$ layer of the slab are presented in Figs.~\ref{fig:fig3} (b) and (c), respectively. 
The Zr $d$ character of bands is color-coded in red. Clearly, for the outermost layer, several bands are observed 
that do not have a counterpart in the electronic structure of the bulk. The most prominent surface state has a dominant Zr $d$ character and crosses $E_F$ twice between the $\Gamma$ and $M$ points. 
Experimentally, an electronic feature assigned to this surface state has been found to strongly hybridize with silicene-derived electronic states 
upon donation of negative charge from adsorbed potassium atoms \cite{Friedlein}, and may be crucial for silicene-substrate interactions even in the pristine case.

\section{The zirconium diboride (0001) surface with single adsorbed silicon atoms}

\begin{figure}[p]
\includegraphics[width=0.90\columnwidth,clip=true,angle=0]{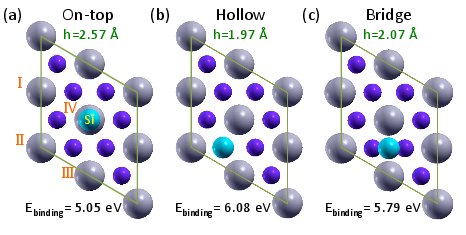}
\caption{\label{fig:fig4}
Individual Si atoms at (a) the on-top, (b) the hollow, and (c) the bridge positions plotted for one Si atom per (2$\times$2) unit cell of the ZrB$_2$(0001) surface. 
The corresponding binding energy per Si atom ($E_{binding}$) and the height ($h$) to its nearest Zr atom are shown as well. The distance to its nearest Zr atom can be found as (a) 2.57~\AA, (b) 2.68~\AA, and (c) 2.61~\AA. 
}
\end{figure}

For the study of the adsorption of Si atoms on the ZrB$_2$(0001) surface, the (2$\times$2) unit cell of ZrB$_2$(0001) 
is considered. This unit cell has been chosen because of its commensurate relationship with the $\sqrt{3}\times\sqrt{3}$ unit cell of reconstructed silicene.
As shown in Figs.~\ref{fig:fig4} (a) to (c), three possible symmetric sites -- the on-top, hollow, and bridge positions -- allow Si atoms to avoid in-plane forces. 
The positions have been relaxed along the out-of-plane direction. For each position, the corresponding heights as given with respect to the uppermost Zr atom and the binding energies are 
defined as (E[ZrB$_2$ slab]+E[two Si atoms]-E[ZrB$_2$ slab + two Si atoms])/2 and indicated in Fig.~\ref{fig:fig4} as well. It is found that the on-top position 
is the least favorable one. 
As understandable from the corresponding coordination numbers, Si atoms prefer the hollow and bridge sites which brings them closer to the Zr layer. 
Obviously, the on-top Si atom that is furthest away from the Zr layer can bind only to a single Zr atom and is thus more weakly interacting which is different from the hollow and bridge sites where three and two Zr atoms are affected, respectively. 

For all positions, the Zr atoms adjust their heights as well such that those closest to the added Si atom adopt a slightly lower position 
with respect to the remaining Zr atoms in the terminating layer. The out-of-plane coordinates are listed in Table~\ref{tab:table1}. 

\begin{table}[h]
\caption{\label{tab:table1} The heights of outermost Zr atoms [c.f. Fig.~\ref{fig:fig4}] measured from the second outermost Zr layer for a single Si atom adsorption on ZrB$_2$(0001) surface. The values without Si atom adsorption are also shown as comparison. The unit is \AA.}
\begin{ruledtabular}
\begin{tabular}{ccccc}
Zr atom &  \multicolumn{4}{c}{Si adsorption site} \\
   & no Si & on-top & hollow & bridge  \\
\hline
I  & 3.48 & 3.54 & 3.56 & 3.59 \\
II & 3.48 & 3.54 & 3.50 & 3.59 \\
III& 3.48 & 3.54 & 3.50 & 3.46 \\
IV & 3.48 & 3.37 & 3.50 & 3.46 \\
\end{tabular}
\end{ruledtabular}
\end{table}

From the information thus obtained, favorable in-plane positions for the Si atoms forming silicene on the ZrB$_2$(0001) surface can be predicted. 
The best candidate would be a structure in which all of the Si atoms reside at hollow sites. 
This is, however, impossible to realize for six Si atoms per (2$\times$2) unit cell since the Si honeycomb lattice prefers a larger lattice constant 
than the boron sub-lattice of ZrB$_2$. On-top sites could also be avoided by locating all six Si atoms in near-bridge sites, that is in positions in between on-top and bridge sites. 
However, the deviation from the perfect bridge position would increase the energy too much 
such that it is preferable to place two Si atoms at hollow sites, three Si atoms at near-bridge sites and one Si atom at an on-top site. 
This simple analysis provides the in-plane structure model of silicene that has independently been derived from experimental data.\cite{Fleurence}

If we assume that Si atoms would adopt the height calculated for individual Si atoms as listed in Figs.~\ref{fig:fig4} (a) to (c), 
the planar-like silicene structure can easily be accommodated on the ZrB$_2$(0001) surface. 
With the protruding Si atom sitting at an on-top position, the other five Si atoms can be located at hollow sites and near-bridge sites, for two and three Si atoms, respectively. 
On the contrary, for the regularly-buckled silicene structure, the height profile does not follow the favorable positions found by this simple analysis 
since the Si atoms sitting on hollow and on-top sites belong to the same sub-lattice. With the restrain of limiting the surface density to exactly six atoms per (2$\times$2) unit cell of the ZrB$_2$(0001), 
the proposed planar-like structure of silicene is indeed calculated to be the ground state, as shown in section VI. 
This points to the dominant role of the local environment of individual silicon atoms for the structural properties of epitaxial silicene. 

Note that with approximately 5 eV, the binding energy per Si atom is quite large. Due to the symmetry breaking caused by the added Si atom and the interactions involved, for all of the three adsorption sites, 
the Zr-$d$-derived surface state splits into bonding and anti-bonding states of which the former is shifted down in energy and the latter becomes unoccupied (band structures not shown). 
With hollow sites showing the strongest interactions, the corresponding Si $p_z$ orbitals hybridize most strongly with Zr $d$ orbitals of the outermost layer.

\section{Epitaxial silicene on the zirconium diboride (0001) surface}

\begin{figure*}[p]
\includegraphics*[width=0.55\columnwidth,clip=true,angle=90]{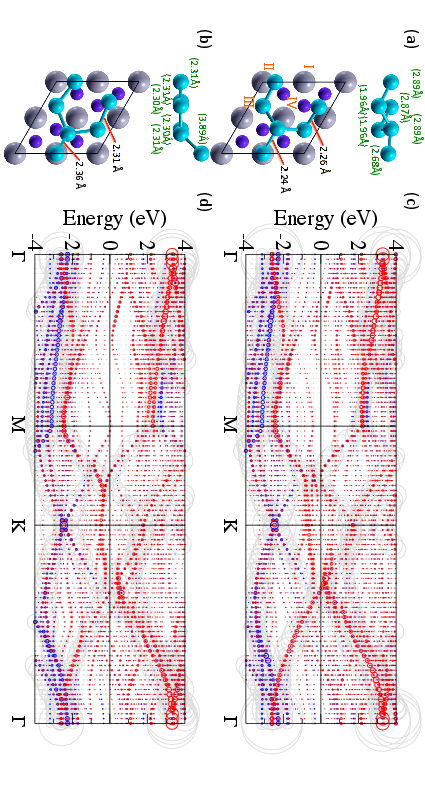}
\caption{\label{fig:fig5}
In-plane and out-of-plane structural parameters of (a) regularly-buckled-like silicene and (b) planar-like silicene on ZrB$_2$(0001). The relative heights of Si atoms to the outermost Zr atom are indicated in brackets.
The heights of the Zr atoms with respect to the second outermost Zr layer are I: 3.63~\AA, II: 3.47~\AA, III: 3.47~\AA, 
and IV: 3.47~\AA~ and I: 3.55~\AA, II: 3.50~\AA, III: 3.50~\AA, and IV: 3.50~\AA. 
(c) and (d): The band structures of the phases shown in (a) and (b).  The Zr $d$ and B $p$ orbital characters are colored in red and blue, respectively. 
Only the weight at the outermost ZrB$_2$ layer is shown in the fat-band representation. The first Brillouin zone of bulk ZrB$_2$ is adopted.
}\end{figure*}

\begin{figure*}[p]
\includegraphics*[width=0.55\columnwidth,clip=true,angle=90]{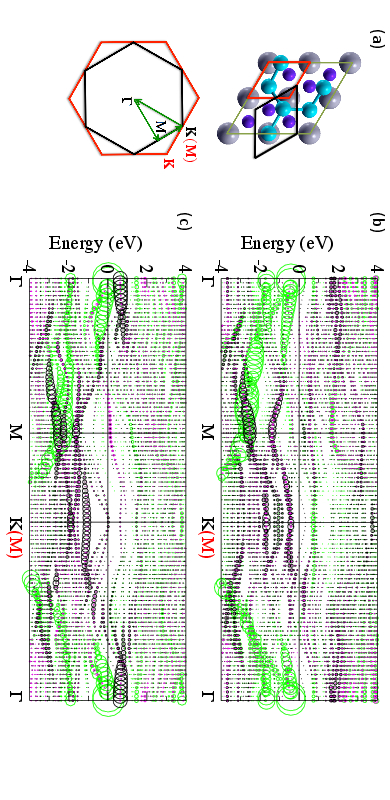}
\caption{\label{fig:fig51}
(a) The (1x1) unit cells of ZrB$_2$ and epitaxial silicene and the corresponding first Brillouin zones. 
The silicon contributions to the band structures of (b) regularly-buckled-like and (c) planar-like silicene on ZrB$_2$(0001), represented in the first Brillouin zone of (1x1) silicene.  
The Si $p_z$ and Si $s$ orbital characters are colored in black and magenta, respectively, and  
the Si $p_x$ and $p_y$ characters are colored in green in the fat-band representation. 
The spectral weight by summing over the orbital contributions of an eigenstate is not shown here.
}
\end{figure*}

In this section, the results for six Si atoms per (2$\times$2)-reconstructed ZrB$_2$(0001) unit cell are reported. 
In the search for possible silicene structures beyond the one predicted in section V, only two minima of the total energy have been found. 
The associated two phases, shown in Figs.~\ref{fig:fig5} (a) and (b), resemble structurally free-standing ones, which are the regularly-buckled and planar-like phases, considered in section III: 
The binding energy per Si atom of these two phases on the diboride surface are 1.14 eV and 1.42 eV, respectively.
In agreement with the predictions for preferential sites for the adsorption, 
planar-like, ($\sqrt{3}\times\sqrt{3}$)-reconstructed silicene is calculated to become the most stable epitaxial structure on ZrB$_2$(0001). 
When placing the planar-like silicene phase onto the surface, the original phonon instability mentioned for free-standing, planar-like silicene in section III,
is balanced by the gain of energy of those atoms such that the planar-like structure becomes stable against out-of-plane vibrations of the Si atoms above hollow sites.
The second, metastable phase, is identified as a deformed version of regularly-buckled silicene 
where atoms of the top sub-lattice are in near-bridge sites and those of the bottom sub-lattice in hollow and on-top sites\cite{Fleurence}. The on-top Si atom resides at a position higher than those on hollow sites.
In the following, we will refer to this structure as the ``regularly-buckled-like'' structure.

For both epitaxial structures, the total binding energy per Si atom defined as (E[ZrB$_2$ slab]+E[two silicene layers]-E[ZrB$_2$ slab + two silicene layers])/12 
is just a little bit larger than 1.0 eV and therefore less than one third of the binding energy calculated for single Si atoms adsorbed on the same surface. 
This suggests that the bonding between Si and Zr atoms is weakened by the strong bonding within the silicene layer. However, 
the bonding of silicene to the diboride surface is still strong enough to significantly impact the electronic structure of the silicene-substrate interface.

\begin{figure}[p]
\includegraphics[width=0.70\columnwidth,clip=true,angle=90]{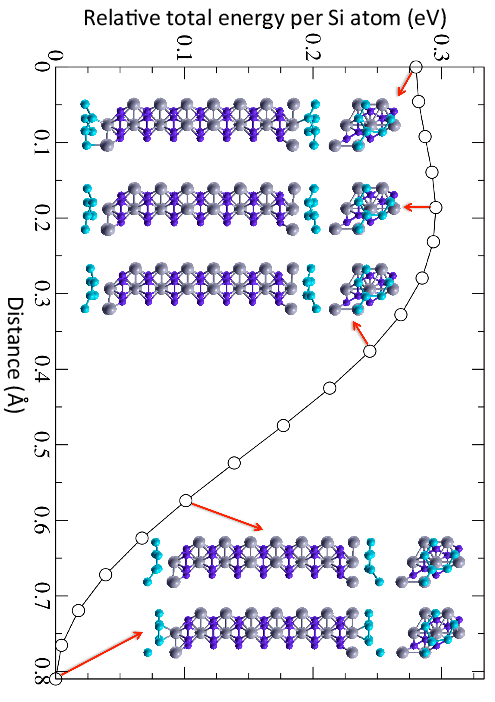}
\caption{\label{fig:fig6}
The energy barrier between the regularly-buckled-like (on the left) and the planar-like (on the right) phases on the diboride surface. The distances of transition images are measured from the regularly-buckled-like configuration. The top view and a side view of selected images are presented. 
}
\end{figure}

The band structures of the regularly-buckled-like and planar-like epitaxial phases are shown in Figs.~\ref{fig:fig5} (c) and (d), respectively.
Along the $\Gamma$-$M$ direction of the ZrB$_2$ (0001) surface, 
hybridization between Zr $d$ and Si $p_z$ orbitals leads to a splitting of the original Zr-derived surface state resulting in the formation of occupied bonding and unoccupied anti-bonding bands
that have parabolic dispersions resembling that of the bare ZrB$_2$(0001) surface. These bands are best identifiable in the planar-like phase. 
For both phases, B $p$-derived electronic states appear to be almost unaffected by the adsorption of silicene.  

While for the discussion of substrate-derived electronic states, the use of the ZrB$_2$ Brillouin zone is appropriate, 
silicene-derived states are best discussed using the silicene (1$\times$1) Brillouin zone. 
The relation between the two is illustrated in Fig.~\ref{fig:fig51} (a). This would then facilitate comparison with the results for free-standing silicene. 
As shown in Figs.~\ref{fig:fig51} (b) and (c), for both phases, at the silicene (1$\times$1) $K_{Si}$ point and in the vicinity of $E_F$, X-shaped dispersing bands are now missing. 
This is very different from the case of free-standing silicene in which cone-like dispersions with $p_z$ character can easily be identified. 
However, for the regularly-buckled-like structure, several bands with partial $p_z$ character are still found below $E_F$ (black color in Fig.~\ref{fig:fig51} (b)).  
For the one closest to the Fermi energy, in particular, the Si $p_z$ contributions are dominant. 
This is consistent with the experimental observation of upward two curved features that approach the Fermi level by up to 250 meV at $K_{Si}$ point \cite{Fleurence} that have been observed recently by angle-resolved photoelectron spectroscopy (ARPES).\cite{Friedlein} 
On the other hand, for the planar-like phase, two upward dispersing bands cut through $E_F$ in the vicinity of the $K_{Si}$ point. 
These bands have only a minor $p_z$ contribution and represent $\sigma$ bands that are folded from the $\Gamma$ point as a consequence of the breaking of the translational symmetry. 
These features do not resemble the features in the ARPES spectra. However, it may be conceived that the parabolic conduction band shifted below the Fermi level following the adsorption of K atoms located half way between the $\Gamma$ and $K$ points of silicene (at the M point of the (2x2)-reconstructed ZrB2(0001) surface) \cite{Friedlein} bares resemblance to the $p_z$-derived conduction band appearing in the calculations.
Instead, a flat $\pi$ band is found at about 1.0 eV below $E_F$ that carries a larger Si $p_z$ weight. 
It is obvious that the changes in the electronic structures occurring in particular for the planar-like phase upon adsorption on the diboride surface are not only consequence of the ($\sqrt{3}\times\sqrt{3}$) 
reconstruction but to a large degree also caused by interactions with the substrate. 

In order to understand if these two epitaxial silicene phases on the ZrB$_2$(0001) surface can be transformed into each other, 
molecular dynamics calculations have been performed as well. 
Regardless of whether the initial structure is close to that of the regularly-buckled-like form or not, the calculations always converge to the planar-like phase  
corresponding to the ground state. This indicates that the energy barrier from the regularly-buckled-like phase to the planar-like phase is low and has been confirmed by the NEB calculation to be about 15 meV per Si atom in contrast to the deep well at the planar-like configuration. 
Selected images of transition states and the total energy along the minimum energy path are plotted in Fig.~\ref{fig:fig6}.  
Here, the total energy of the planar-like phase has been set to zero. 

\section{Discussion and summary}

Given that a purely $sp^2$-hybridized, planar free-standing silicene is unstable, 
it is of great interest to understand the mechanisms that stabilize Si honeycomb lattice configurations when placed epitaxially on surfaces. 
Along this line, we have investigated planar-like and regularly-buckled-like ($\sqrt{3}\times\sqrt{3}$)-reconstructed silicene phases whose occurance has been predicted for epitaxial silicene systems.  
Additionally, these phases have been compared to their respective free-standing, non-reconstructed forms. 
For a large range of the in-plane lattice constant, free-standing, ($\sqrt{3}\times\sqrt{3}$)-reconstructed, 
planar-like silicene is less stable than the regularly buckled form in terms of the total energy. 
Since it is characterized by one Si atom protruding out of the plane defined by the other five Si atoms of the unit cell, the electronic structure 
is found to reflect the mixed $sp^2$/$sp^3$ bonding nature of this phase. Due to the broken symmetry of the reconstructed surface, the degeneracy of electronic states at the Dirac point of unreconstructed silicene is lifted. 

However, with restraining the surface density to exactly six atoms per (2$\times$2) unit cell, the planar-like phase has been found to be the ground state on 
the Zr-terminated ZrB$_2$(0001) surface. This is because the positions of individual Si atoms adopted in this phase are stabilized by interactions with Zr $d$ orbitals.
However, as outlined recently, the experimental data are more consistent with the presence of the metastable phase resembling the regularly-buckled form of free-standing silicene \cite{Fleurence}
and with rather weak interactions at the silicene-diboride interface \cite{Friedlein}. 
In particular, the height profile of the Si honeycomb structure measured by STM\cite{Fleurence} does not show any corrugations higher than 0.5~\AA. Photoelectron diffraction data 
are consistent with a structure model in which two of the six Si atoms of the ($\sqrt{3}\times\sqrt{3}$) unit cell in the so-called ``A'' positions are surrounded by three ``B'' atoms 
at a higher position. Recent ARPES spectra indicate that the parabolic diboride surface state and a silicene-derived band 
hybridize only after electron donation to the surface following the adsorption of potassium atoms \cite{Friedlein}.

On the other hand, epitaxial silicene on the ZrB$_2$(0001) surface is characterized by the spontaneous formation of stress domains indicating 
the presence of long range interactions within the two-dimensional layer of the Si ad-atoms.\cite{Fleurence} 
Having the discussion of the theoretical results presented in this paper in mind, the available experimental data suggest 
that epitaxial strain weakens the interactions with the substrate such that the stability of possible silicene phases on substrates 
can be reversed. In order to comprehend the impact of epitaxial strain, it may then be necessary to perform large-scale calculations 
covering an extended parameter set which correctly include the effect of interactions between the experimentally observed domains. 
Nevertheless, it is interesting to note that the ground state of silicene on ZrB$_2$ revealed in our DFT calculations, the ($\sqrt{3}\times\sqrt{3}$)-reconstructed planar-like phase, 
has also been calculated to form on the Ag(111) surface \cite{Chen2}. Compared to that of free-standing silicene, the phase diagram of epitaxial silicene may be thus much richer.
 
\bibliography{refs}
\end{document}